\documentclass[11pt]{article}
\usepackage{moriond,epsfig}

\def\Journal#1#2#3#4{{#1} {\bf #2}, #3 (#4)}

\def\PRL{\em Phys. Rev. Lett.}
\def\PRD{{\em Phys. Rev.} D}

\def\be{\begin{equation}}
\def\ee{\end{equation}}
\def\bea{\begin{eqnarray}}
\def\eea{\end{eqnarray}}

\input myDefinitions.tex

\begin{document}
\vspace*{4cm}
\title{NEW QUARKONIUM RESULTS FROM THE BABAR EXPERIMENT}

\author{N. ARNAUD, representing the \babar~Collaboration}

\address{Laboratoire de L'Acc\'{e}l\'{e}rateur Lin\'{e}aire, IN2P3/CNRS et Universit\'{e} Paris XI,\\F-91898 Orsay Cedex, France}

\maketitle\abstracts{
New \babar~results on \B mesons and quarkonia are presented: an analysis of \bxkch and \bxkneu decays with \xpsipipi, a precise measurement of the \B mass difference $\Delta m_\B = m(\Bz)-m(\Bp)$ and a study of hadronic transition between $\Upsilon$ mesons.
}

\section{Introduction}

With the end of the data taking phase, the \babar~collaboration is now entering a intensive phase of analysis aiming at exploiting the huge \FourS data sample collected. The studies presented here benefit from the large \bbbar~and \ccbar~cross-sections and provide several new or improved measurements in the quarkonium area. They all have been optimized keeping the signal regions blind. In addition, a new measurement of the \B mass difference is presented.

\section{Study of the decays \bxk with \xpsipipi}

The \Xz~state, discovered\cite{BelleDiscovery} by the Belle collaboration, does not match any predicted charmonium state: many theoretical ideas have been proposed to explain its existence. The mass\cite{pdg} of this narrow resonance, $\Gamma_X < 2.3 \mevcc$ at 90\% confidence level (C.L.), is barely above the $\bar{D^0}D^{*0}$ threshold: $m_X = 3871.4 \pm 0.6 \mevcc$. Although its decay to this state has been observed\cite{belled0d0bar,babard0d0bar}, the measured mass is significantly higher (about 3\mevcc); whether this discrepancy is the sign of two distinct states or a threshold effect is still unclear\cite{dz}. The \Xz~quantum numbers are not known although its parity should be positive\cite{BabarXjpsigamma} and angular analyzes\cite{CDFXAngular} favour $J^{PC} = 1^{++} \ {\rm or}\ 2^{-+}$.

\begin{figure}[here!]
\begin{center}
\begin{tabular}{cc}
\includegraphics[width=0.35\textwidth]{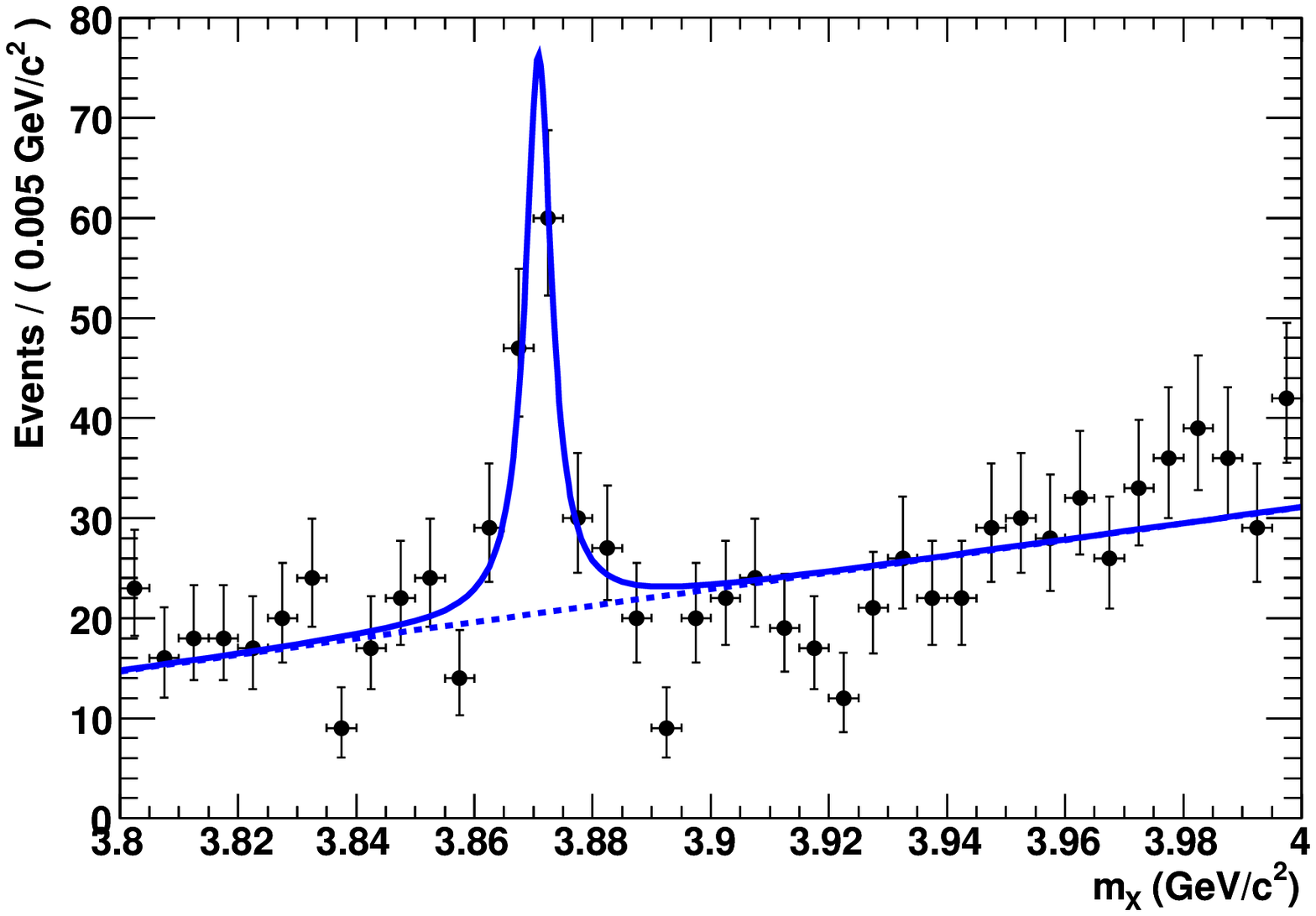}&
\includegraphics[width=0.35\textwidth]{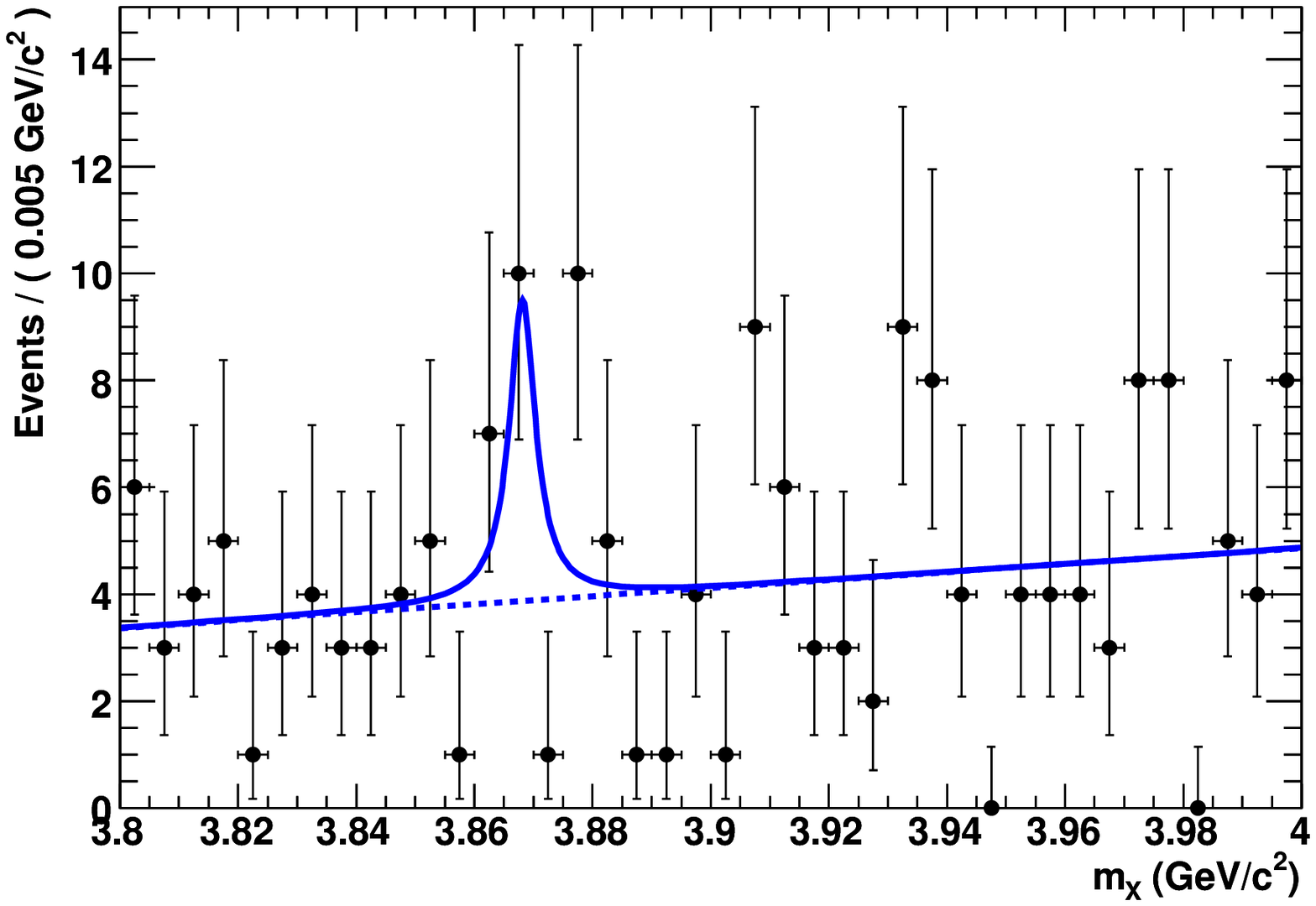}
\end{tabular}
\caption{Fits to the $m(\jpsi\pip\pim)$ data distributions of (left) $\Bp\to\Xz\Kp$ and (right) $\Bz\to\Xz\KS$ candidates. The dashed (solid) line represents the background (sum of background plus signal) probability density function. The statistical significances w.r.t. the null signal hypothesis are $8.6\sigma$ and $2.3\sigma$ respectively.
\label{fig:X3872}}
\end{center}
\end{figure}

The decays \bxk with \xpsipipi are studied\cite{philippejack} on the full \babar~dataset (413~\invfb). Signal events are discriminated from background using two kinematical variables: the energy difference $\DeltaE = E_B - \sqrt{s}/2$ and the beam-energy substituted mass $\mes = \sqrt{s/4-(\vec{p}_B)^2}$ where $(E_B,\vec{p}_B)$ is the \B 4-momentum vector in the \FourS rest frame. $\DeltaE$ (\mes) peaks at 0 ($m_B$) for signal with a resolution of a few tens of (a few) \mev. In addition, event-shape variables reject random combinations of particles coming from continuum (\qqbar) events.

The numbers of signal events for the charged and neutral modes are extracted by an unbinned maximum likelihood (UML) fit of the \Xz reconstructed mass in the \mes~signal region. $\B \to \psitwos K$ with \psitwospsipipi decays are used as a control sample (same final state, close mass and very narrow state) to validate the fitting procedure and estimate biases. For both modes, the dominant systematics come from the background modeling and the secondary branching fractions. For the neutral mode, the fit convergence is ensured by fixing the resonance width to the value obtained from the charged mode fit, which induces the largest systematics. The results are:

\begin{equation}
   \frac{\BR ( \Bz\to\Xz\Kz ) }{ \BR ( \Bp\to\Xz\Kp ) } 
   = \frac{(3.5\pm 1.9 \pm 0.4) \times 10^{-6}}{(8.4\pm 1.5 \pm 0.7) \times 10^{-6}}
   = 0.41\pm0.24 \pm0.05
   < 0.73 \mathrm{\; at \, 90\% \, C.L.}
\end{equation}

Still using the \psitwos~control sample, the mass difference of the \Xz states produced in \Bz and \Bp decays is found to be $2.7\pm 1.6\pm 0.4\mevcc$. These results are consistent with either the molecular\cite{swavemolecule} or diquark-antidiquark\cite{diquark} model within two standard deviations. Finally, an updated upper limit of the \Xz natural width is set to $\Gamma_{\rm X}<3.3~\mevcc$ ($90\%$, C.L.).

\section{Measurement of the \B mass difference}

The neutral-to-charged mass difference of \B mesons probes Coulomb contributions to their quark structures and is needed to determine \Bp~and \Bz~decay fractions at \B-factories. To improve its accuracy (much worse\cite{pdg} than for the \pion, \kaon and $D$ mesons), a method based on the analysis of momentum spectra in the decays \bpsik has been applied\cite{bmassdiff} on a sample of 230 million \BB~events. In the center-of-mass frame (related quantities labeled with a $^*$), one has:

\begin{equation}
\Delta m_\B \equiv m(\Bz)-m(\Bp) = -\left[ p^*(\Bz) - p^*(\Bp) \right] \;
              \frac{p^*(\Bz) + p^*(\Bp)}{m(\Bz) + m(\Bp)}
\label{eq:bmassdiff}
\end{equation}

As the \BB~threshold is very close to the \FourS~mass, \B~mesons have low $p^*$ which makes the right-most factor of Eq.~\ref{eq:bmassdiff} almost constant. Hence, $\Delta m_\B$ can be extracted from the mean values of the $p^*$ spectra. The resolution depends on the accelerator energy spread and on the reconstruction whose resolutions are about 40\mevc~and 15\mevc~respectively for the clean selected decay modes: $\Bz\to\jpsi\Kp\pim$ and $\Bp\to\jpsi\Kp$, where $\jpsi\to\mumu$ or \epem.

The size of a possible bias induced by the mean-$p^*$ method has been estimated by MC simulations. The influence of the beam energy smearing and of the \FourS lineshape is studied using a simulation without detector. Varying the parameters\cite{babar4slineshape} and choosing $\Delta m_\B=0.3$ or 0.4\mevcc, the measured mass differences are within 2\% of the input values. The detector contribution to the bias is computed using a full MC simulation of generic \BB~decays and leads to the dominant systematics. Other important systematics errors are coming from the choice of the fit functions, both for the background and the signal species.

\jpsi~decays to \mumu~and \epem~are dealt with separately to account for bremsstrahlung in the second channel. The selections are based on \babar~particle-identification selectors, event-shape variables and $\DeltaE$. The \Bz~and \Bp~samples are fitted simultaneously as the common effect of the PEP-II energy spread dominates the $p^*$ distribution shapes. The two $p^*$ mean differences are consistent and can then be combined. Adding systematics errors, one finally gets:

\begin{equation}
\Delta p^* = (-5.5\pm 0.8\pm 0.5)\mevc \Rightarrow
\Delta m_\B = (+0.33\pm 0.05\pm 0.03)\mevcc
\end{equation}

The measured \B~mass difference is compatible with the current world average\cite{pdg} with errors reduced by a factor 4. Consequently, the significance of $\Delta m_\B$ being positive exceeds the $5\sigma$ level. Using this measurement to compute the associated contribution to the ratio of \FourS decay rates to charged and neutral \B~pairs gives a value compatible with the experimental results\cite{pdg}: unknown Coulomb contributions to the quark structure\cite{theory} are not visible yet.

\section{Hadronic transitions between $\Upsilon$ states}

The \bbbar~system offers unique opportunities to study hadronic transitions between $\Upsilon$ states with 5 known (4 allowed but not observed so far) transitions with 2 charged pions (1 $\eta$ meson) in the final state. Experimental results allow to test the QCD Multipole Expansion model\cite{qcdme}.

\FourS hadronic transitions are searched in a large data sample: 347~\invfb recorded at the resonance peak ('onpeak') and 37~\invfb taken 40\mev below ('offpeak' data used as control sample). \TwoS and \ThreeS transition studies are based on onpeak data with initial state radiation. $\NS \to \mumu$ or \epem are processed separately to account for trigger efficiency differences and $e^-$ bremsstrahlung. For $\eta$ channels, decays to $\pipi\piz$ are reconstructed: the trigger efficiency is much larger than for the \etagg mode and all final states have the same charged particles, which cancels out several systematics for partial width ratios. Signal events are identified by requiring compatibility between the invariant lepton pair mass $M_{ll}$ and the known mass of a \NS resonance. For $\pipi$ channels, the invariant mass difference $\Delta M=M_{\pi\pi ll}-M_{ll}$ between the final state and the leptons must also match the difference between the masses of the initial and final $\Upsilon$ states. For decays involving an $\eta$ meson, the mass difference takes into account the additional constraint: $\Delta M_\eta=M_{3\pi ll}-M_{ll}-M_{3\pi}$, with $M_{3\pi}$ being the mass of the $\eta$ candidate.

\begin{figure}[here!]
\begin{center}
\begin{tabular}{cc}
\includegraphics[width=0.25\textwidth]{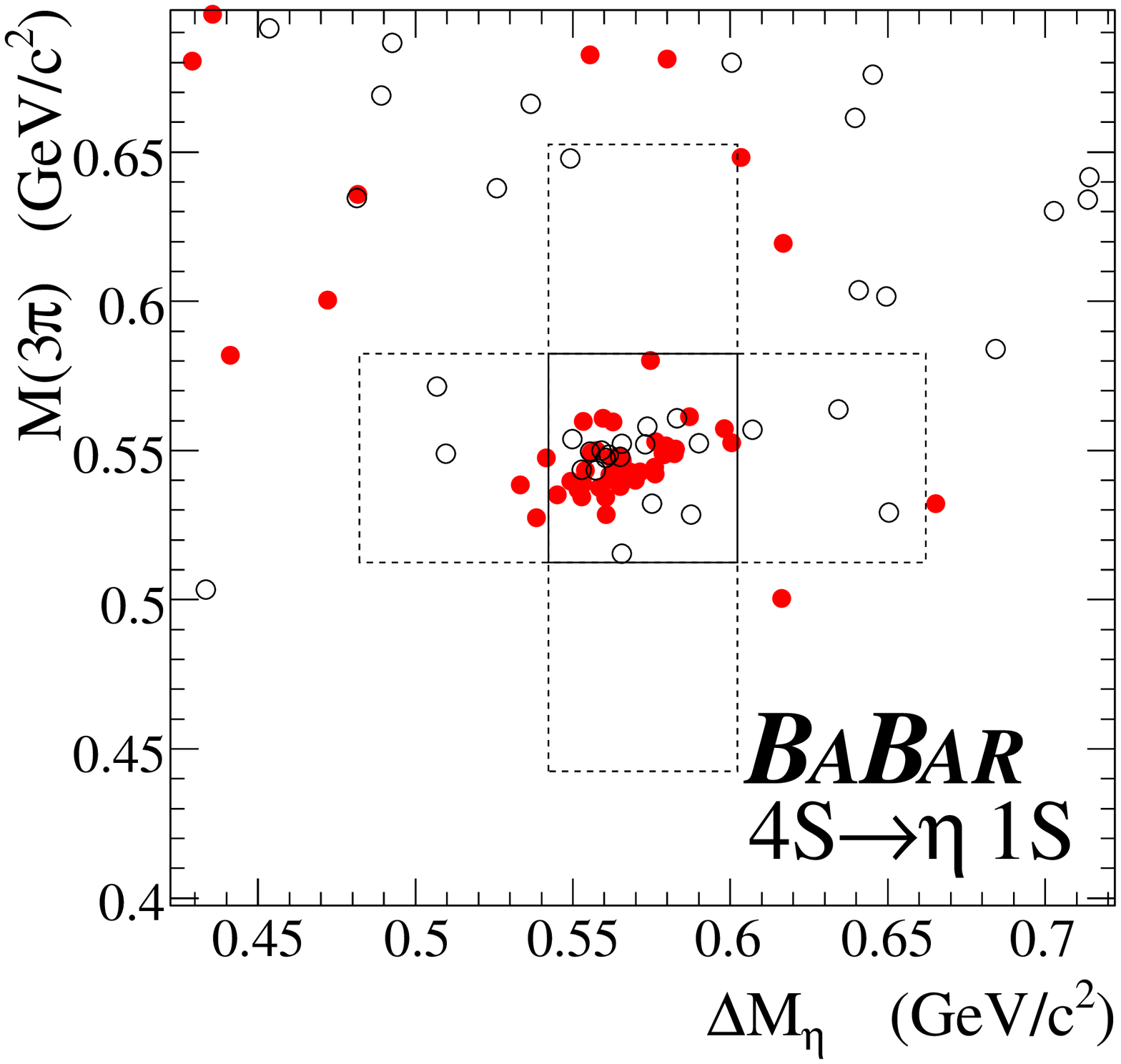}&
\includegraphics[width=0.25\textwidth]{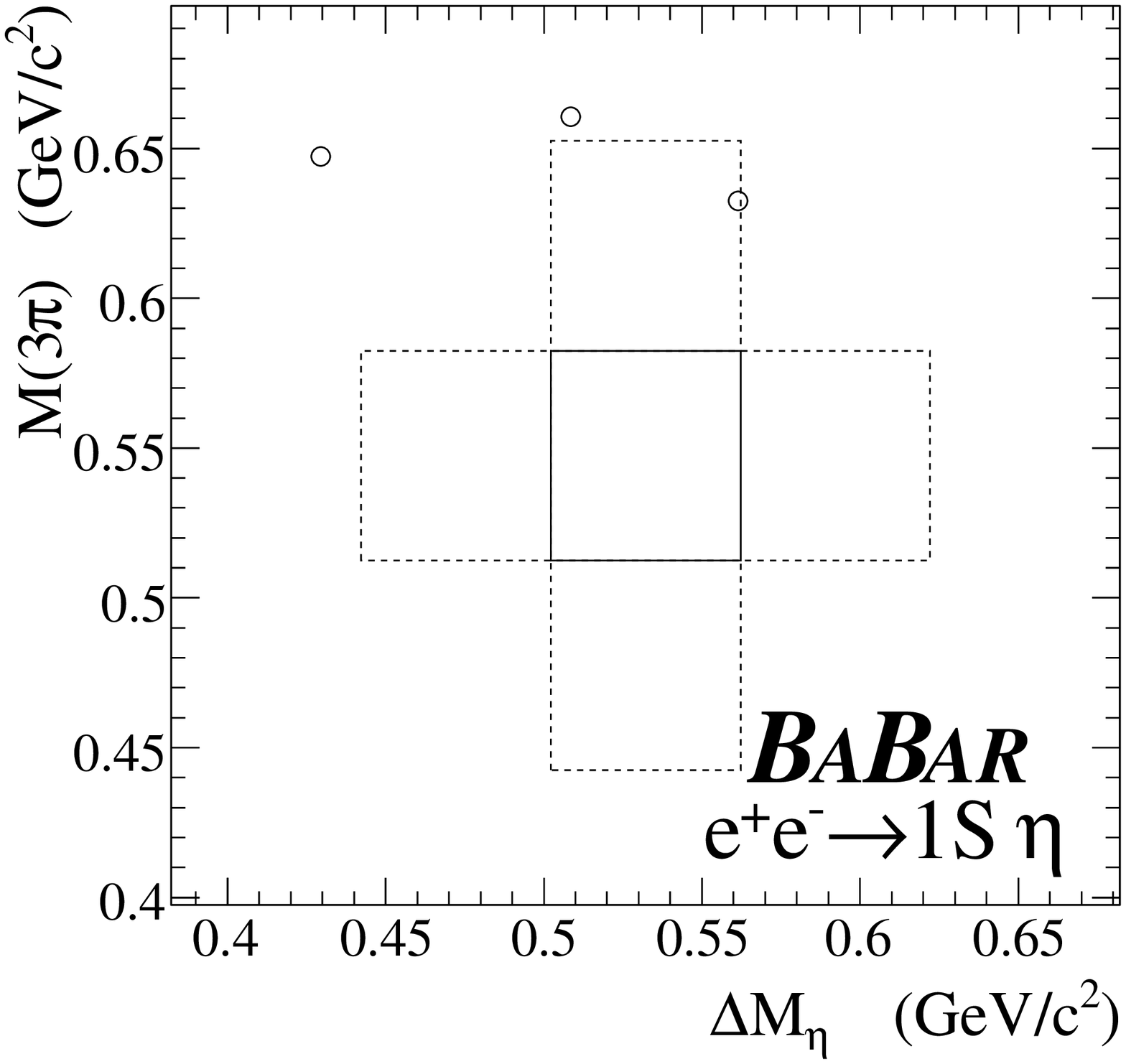} \\
\includegraphics[width=0.25\textwidth]{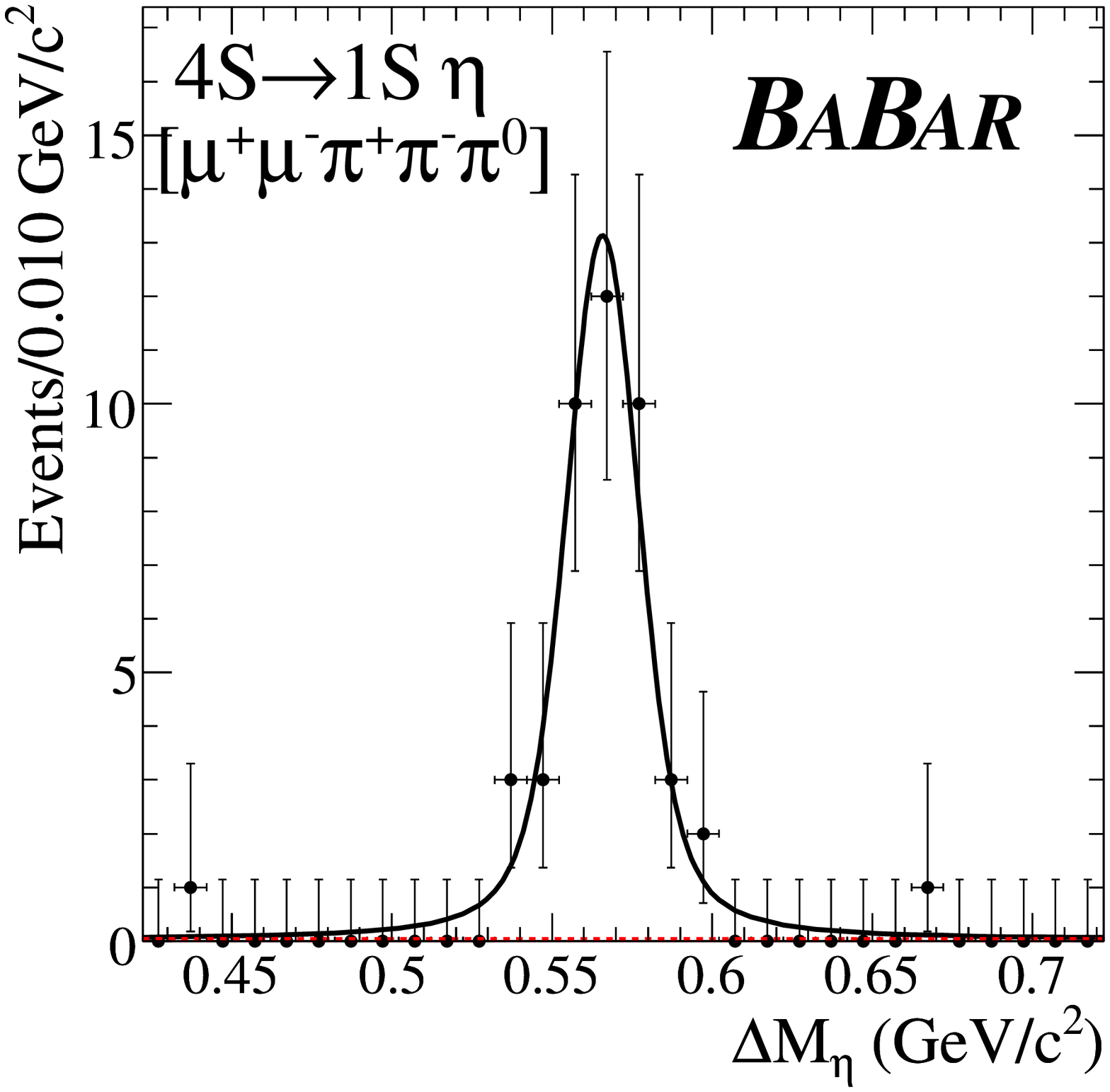}&
\includegraphics[width=0.25\textwidth]{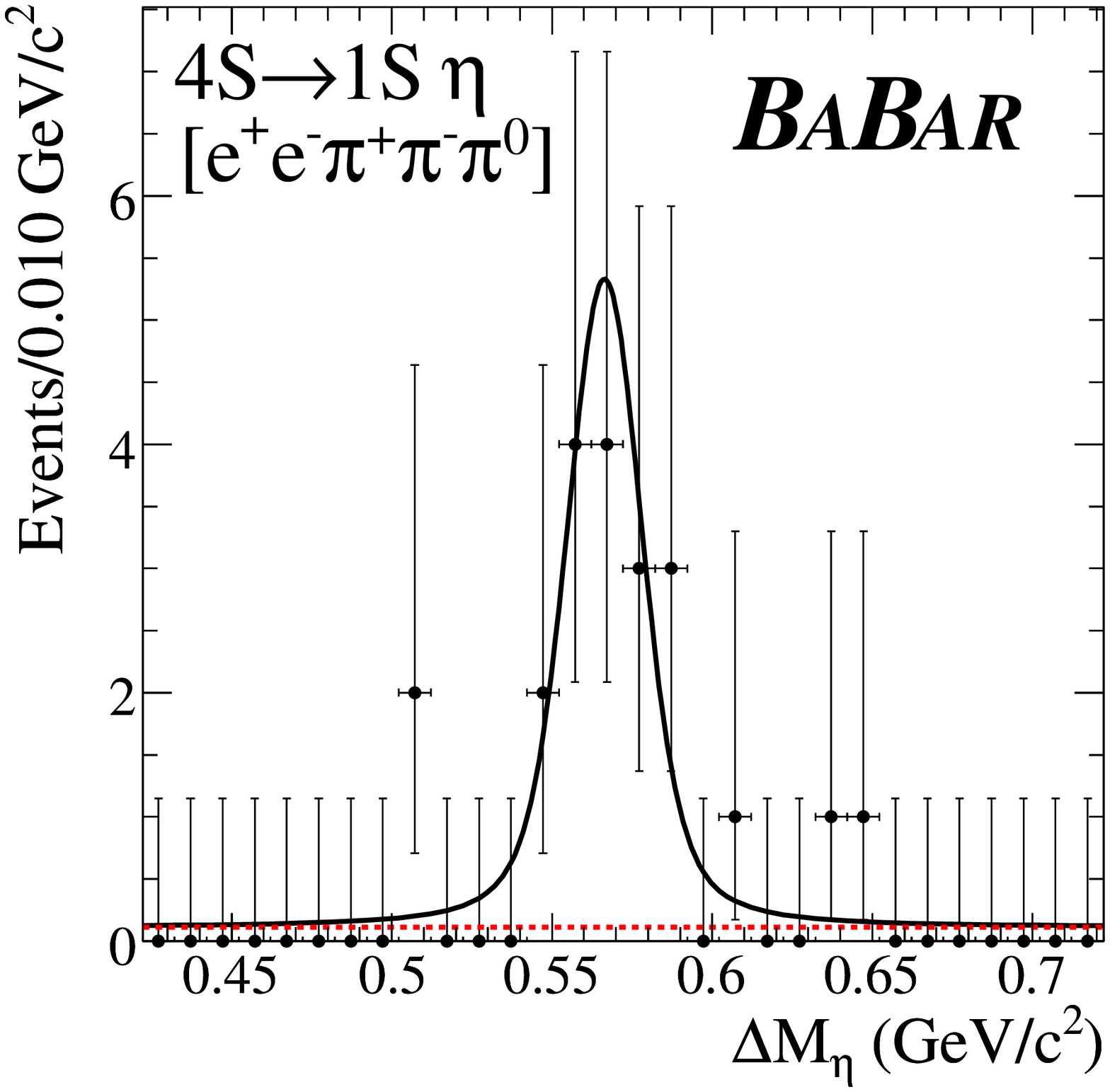}
\end{tabular}
\caption{Top row: $\FourS \to \OneS \eta$ event distributions in the $(\Delta M_\eta$;$M_{3\pi})$ plane for onpeak (left) and offpeak (right) data samples. Open circles (dots) show $\Upsilon$ decays to \epem (\mumu). Solid (dashed) lines delimit the signal (background) box(es). Bottom row: fits to the signal candidates with $\OneS \to \mumu$ (left) and $\epem$ (right). The statistical significances w.r.t. the null signal hypothesis are $11\sigma$ and $6.2\sigma$ respectively.
\label{fig:upsilons}}
\end{center}
\end{figure}

The efficiency-corrected signal yields for the known $\MS \to \NS\pipi$ transitions are determined without any assumption on the decay angular distribution. Candidates are classified using two variables: the $\pi\pi$ invariant mass and the cosine of the helicity angle between the $\pip$ direction in the $\pi\pi$ rest frame and the $\pi\pi$ direction in the $\Upsilon$ rest frame. In each bin, the signal yield is computed using an extended UML fit to the $\Delta M$ distribution. Then, the global yield is computed by summing these numbers weighted by the inverse of the MC signal bin efficiency. Improved measurements of partial width ratios are for instance $\GG{\FourS}{\pipi\TwoS}/\GG{\FourS}{\pipi\OneS}=1.16\pm 0.16 \pm 0.14$ and $\GG{\ThreeS}{\pipi\TwoS}/\GG{\ThreeS}{\pipi\OneS}=0.577\pm 0.026 \pm 0.060$.

For the $\MS \to \NS \eta$ transitions, events are studied in the plane $(\Delta M_\eta$;$M_{3\pi})$ where signal and sideband (used for background extrapolation) boxes, are defined for each decay. No signal excess is found for $\TwoS \to \OneS \eta$ and $\ThreeS \to \OneS \eta$ but 56 (0) $\FourS \to \OneS \eta$ candidates are observed in the onpeak (offpeak) sample. The probability that they come from continuum is 0.3\%. The numbers of signal events for the $\mumu$ and $\epem$ $\Upsilon$ decay modes are extracted via extended UML fits in the $(\Delta M_\eta)$ variable. Including systematics, the results are

\begin{eqnarray}
{\cal B}(\FourS\to\eta\OneS) &=& (1.96\pm0.06\pm0.09)\times 10^{-4} \\
\DISP\frac{\GG{\FourS}{\eta\OneS}}{\GG{\FourS}{\pipi\OneS}} &=& 2.41\pm0.40\pm0.12
\end{eqnarray}
the latter ratio being surprisingly large with respect to others (although there is no prediction for this particular decay mode): the same analysis gets 90\% C.L. upper limits of $5.2 \times 10^{-3}$ and $1.9 \times 10^{-2}$ for the $\TwoS \to \OneS \eta$ and $\ThreeS \to \OneS \eta$ transitions respectively. 

\section{Conclusion}

New and improved measurements in the quarkonium area have been presented by the \babar~collaboration. They cover a wide range of topics: the new \Xz state, the \B-mass difference and the hadronic transitions between $\Upsilon$ states. 


\end{document}